\documentclass[12pt,prd,aps,amssymb,amsmath,tightenlines,showpacs]{revtex4}
\usepackage{graphicx,epsfig,color}

\def\cP{\mathcal P}

\def\cT{\mathcal T}
\def\cA{\mathcal A}
\begin{document}

\topmargin=0.0cm

\title{Critical behavior of the $\cP\cT$-symmetric $i\phi^3$ quantum field
theory}

\author{Carl M. Bender$^a$}\email{cmb@wustl.edu}
\author{V. Branchina$^b$}\email{branchina@ct.infn.it}
\author{Emanuele Messina$^b$}\email{emanuele.messina@ct.infn.it}

\affiliation{${}^a$Department of Physics, Washington University, St. Louis, MO
63130, USA\\
${}^b$Department of Physics, University of Catania and INFN, Sezione di
Catania, Via Santa Sofia 64, I-95123 Catania, Italy}

\begin{abstract}
It was shown recently that a $\cP\cT$-symmetric $i\phi^3$ quantum field theory
in $6-\epsilon$ dimensions possesses a nontrivial fixed point. The critical
behavior of this theory around the fixed point is examined and it is shown
that the corresponding phase transition is related to the existence of a
nontrivial solution of the gap equation. The theory is studied first in the
mean-field approximation and the critical exponents are calculated. Then, it
is examined beyond the mean-field approximation by using renormalization-group
techniques, and the critical exponents for $6-\epsilon$ dimensions are
calculated to order $\epsilon$. It is shown that because of its stability the
$\cP\cT$-symmetric $i\phi^3$ theory has a higher predictive power than the
conventional $\phi^3$ theory. A comparison of the $i\phi^3$ model with the
Lee-Yang model is given.
\end{abstract}

\date{\today}
\maketitle

\section{Introduction}
\label{s1}

The study of $\cP\cT$-symmetric quantum theory \cite{R1} was originally
motivated by the discovery that the eigenvalues of the quantum-mechanical
Hamiltonian $H=p^2+ix^3$ are real, positive, and discrete \cite{R2}. This work
naturally led to a number of studies of the properties of a scalar quantum field
theory having an imaginary cubic self-interaction term \cite{R3}. A very recent
study of the renormalization group (RG) equations for the $\cP\cT$-symmetric
$ig\phi^3$ quantum field theory in $d=6-\epsilon$ dimensions shows the existence
of a nontrivial fixed point \cite{R4}. This allows for a nonperturbative
renormalization of the theory and suggests that the theory undergoes a
continuous phase transition.

In the present paper we study this transition in detail; that is, we examine the
critical behavior of the theory. We will see that such a transition is
associated with the existence of a nontrivial solution to the gap equation at a
critical value $m^2_c$ of the bare mass $m^2$. The correlation length (the
inverse of the renormalized scalar mass) diverges at $m^2=m^2_c$.

In Sec.~\ref{s2} we study this transition within the framework of the mean-field
approximation and in Sec.~\ref{s3} we obtain the critical behavior of the theory
by calculating the critical exponents. Next, in Sec.~\ref{s4} we compare the
conventional and the $\cP\cT$-symmetric $\phi^3$ theories and analyze the
relation between their renormalization and stability properties. We show that
compared with the conventional $\phi^3$ model, the $\cP\cT$-symmetric $i\phi^3$
theory exhibits new and interesting features. Remarkably, the $\cP\cT$-symmetric
theory has a higher predictive power than the $\phi^3$ theory. This is because
its critical behavior (and therefore its renormalization properties) are
governed by one parameter less than the conventional theory. We will show that
this property is related to the different stability properties of the two
theories. 

Section~\ref{s5} goes beyond mean-field analysis. We study the theory near $d=6$
dimensions and calculate the critical exponents up to O$(\epsilon)$. For $d<6$,
the fluctuations around the mean-field configuration become important and the
analysis of the critical behavior requires the use of RG techniques. With the
help of the hyperscaling relations, the critical exponents are calculated for
$d=6-\epsilon$. Some conclusions are given in Sec.~\ref{s6}.

\section{Mean-field analysis}
\label{s2}
In Ref.~\cite{R4} a nontrivial fixed point of the RG equations of the $\cP
\cT$-symmetric $ig\phi^3$ quantum field theory in $d=6-\epsilon$ dimensions was
found. The existence of such a fixed point suggested the onset of a continuous
transition. In this section we study this transition by performing a mean-field
analysis.

The partition function $Z[h]$ of the quantum field theory in $d$ dimensions is
given by
\begin{equation}
Z[h]=\int\mathcal{D}\phi\,e^{-S[\phi]-i\int d^dx\,h\phi},
\label{e1}
\end{equation}
where $h$ is an external field. The action is $S[\phi]=\int d^d x\left[(
\partial_\mu\phi)^2/2+V(\phi)\right]$ and the potential is $V(\phi)=m^2\phi^2/2
+ig\phi^3/6$.

We consider the mean-field approximation to $Z[h]$, first searching for a
constant-field saddle point and then performing a semiclassical expansion around
this configuration. [The exponential in (\ref{e1}) contains an implicit factor
of $1/\hbar$, and in the semiclassical approximation $\hbar$ is treated as
small, which justifies the use of steepest-descent asymptotic techniques
\cite{R5}.] Without loss of generality, we assume that $g>0$.

The constant-field saddle points are given by the gap equation
\begin{eqnarray}
m^2\bar\phi+ig\bar\phi^2/2=-ih.
\label{e2}
\end{eqnarray}
When $h=0$, (\ref{e2}) becomes
\begin{eqnarray}
\bar\phi(m^2+ig\bar\phi/2)=0,
\label{e3}
\end{eqnarray}
which has the two solutions $\bar\phi_1=0$ and $\bar\phi_2=2im^2/g$. We must
determine which of the saddle points, $\bar\phi_1$ or $\bar\phi_2$, contributes
to the small-$\hbar$ asymptotic behavior of the path integral (\ref{e1}).

\subsection{Asymptotic analysis for $d=0$}
Let us first examine the problem for the simple case $d=0$. The second
derivative of the exponential in (\ref{e1}) at $h=0$ is $-m^2-ig\phi$. Thus, at
$\phi=\bar\phi_1$ the second derivative is $-m^2$, and at $\phi=\bar\phi_2$ the
second derivative is $m^2$. The sign of the second derivative of the potential
determines the directions of the steepest-descent paths (constant-phase
contours) in the neighborhood of the saddle points. Thus, if $m^2$ is positive,
the {\it down} directions from $\bar\phi_1$ are horizontal (parallel to the
${\rm Re}\,\phi$ axis) and the {\it down} directions from $\bar\phi_2$ are
vertical (parallel to the ${\rm Im}\,\phi$ axis). However, if $m^2$ is negative,
the directions are reversed; the {\it down} directions from $\bar\phi_2$ are
horizontal and the {\it down} directions from $\bar\phi_1$ are vertical.

It is necessary to deform the original integration path in (\ref{e1}), which
lies on the real axis, into a constant-phase contour that passes through the
appropriate saddle point. If we let $\phi=u+iv$ and note that the phase (the
imaginary part of the exponent) at the saddle points vanishes, we obtain a cubic
algebraic equation for the constant-phase contour: $u\left(gv^2/2-m^2v-gu^6/6
\right)=0$. Thus, near $\infty$ the steepest-descent contours asymptote to the
angles $\pi/2$, $-\pi/6$, and $-5\pi/6$ and the steepest-ascent paths asymptote
to the angles $-\pi/2$, $\pi/6$, and $5\pi/6$. Thus, the {\it uniquely
determined} steepest-descent contour is a $\cP\cT$-symmetric
(left-right-symmetric) path that terminates at the angles $-\pi/6$ and $-5
\pi/6$.

Hence, the crucial observation is this: When $m^2>0$, $\bar\phi_2$ lies above
$\bar\phi_1$ on the ${\rm Im}\,\phi$ axis and the steepest-descent path passes
through $\bar\phi_1$ (and not $\bar\phi_2$). However, when $m^2<0$, $\bar\phi_2$
lies below $\bar\phi_1$ on the ${\rm Im}\,\phi$ axis and the steepest-descent
path passes through $\bar\phi_2$ (and not $\bar\phi_1$). A phase transition
occurs at $m^2=0$ where the saddle points coincide.

\subsection{Asymptotic analysis for $d>0$}
When $d>0$, we consider the Hessian matrix $D^{-1}(q,p)=\Delta^{-1}(q^2)
\delta^d(p+q)$ in Fourier space, where
$$\Delta^{-1}(q^2)=\int d^d x d^d y\,e^{-i q(x-y)}\left.\frac{\delta^2S}
{\delta\phi(x)\delta\phi(y)}\right|_{\bar\phi_i}=q^2+m^2+ig\bar\phi_i.$$
This is just the inverse tree-level correlator when $\bar\phi_i$ is the vacuum.
For $m^2\geq0$, $D^{-1}(q,p)$ is positive definite when evaluated at $\bar\phi_1
=0$. In this case we find that
\begin{equation}
\Delta^{-1}(q^2)=q^2+m^2.
\label{e4}
\end{equation}
On the other hand, for $m^2<0$, $D^{-1}(q,p)$ is positive definite when
calculated at $\bar\phi_2=2im^2/g$, and we get
\begin{equation}
\Delta^{-1}(q^2)=q^2+m^2-2m^2=q^2-m^2.
\label{e5} 
\end{equation}
These results imply that there is a phase transition in the Euclidean $\cP
\cT$-symmetric $ig\phi^3$ theory, with the two phases being determined by the
order parameter $\bar\phi$, where $\bar\phi_1=0$ and $\bar\phi_2=2im^2/g$.

We proceed to identify the {\it relevant couplings}; in general, these are the
couplings that control the phase transition. In the ferromagnetic case, the
relevant couplings are the {\it temperature} and the {\it external magnetic
field} \cite{R6}. Thus, we look for the two couplings that in our case play the
role that the temperature and the external magnetic field play in the
ferromagnetic case. We begin by defining the {\it temperature} $T$ as
\begin{equation}
T\equiv 2m^2/g,
\label{e6}
\end{equation}
where the coefficient $2/g$ in (\ref{e6}) is chosen for later convenience
\cite{R7}. The critical temperature $T_c$ is obtained with the help of
(\ref{e3}) for the nontrivial saddle point (that is, the solution to $m^2+ig\bar
\phi/2=0$). We seek the limiting value of $T$ for which $\bar\phi$ vanishes, and
in our case we have $T_c=0$. Therefore, the reduced temperature is given by
\begin{eqnarray}
\tau=T-T_c=2m^2/g.
\label{e7}
\end{eqnarray}
The {\it external magnetic field} (the conjugate field) is just the parameter
$h$.

From (\ref{e4}) and (\ref{e5}) we see that the excitations above the trivial
vacuum $\bar\phi_1=0$ and those above the nontrivial one $\bar\phi_2=2im^2/g$
have the same mass, $M=|m^2|^{1/2}$, which goes to zero as we approach the
critical point. Moreover, as we move continuously from positive to negative
values of $m^2$, $\bar\phi=2im^2/g$ moves continuously down the
negative-imaginary axis starting from $\bar\phi=0$ (when $m^2\geq0$). The
presence of such a divergent correlation length (the inverse of the scalar mass
$M=0$) and of a continuously varying order parameter $\bar\phi$ are the
indications of a continuous (second-order) phase transition. In RG language,
this is due to the existence of the nontrivial fixed point found in
Ref.~\cite{R4}.

We have shown that at least within the framework of the mean-field approximation
used in this section, the fixed point found in Ref.~\cite{R4} governs the
transition from the $\bar\phi_1=0$ phase to the $\bar\phi_2=2im^2/g$ phase. In
Sec.~\ref{s3} we calculate the critical exponents that govern the behavior of
the theory in the critical region in terms of the parameters $h$ and $\tau$.

\section{Critical exponents in the mean-field approximation}
\label{s3}
In this section we give a quantitative description of the critical behavior of
the model for $d>6$ dimensions. We define and calculate the critical exponents
in the mean-field approximation by following the standard terminology used for
Ising-like models. With the help of (\ref{e7}) we write the saddle-point
equation (\ref{e2}) for $h\neq0$ as 
\begin{equation}
\tau+i\bar\phi=-2ih/\left(g\bar\phi\right),
\label{e8}
\end{equation}
which has the typical form of a gap equation. From (\ref{e8}) we can immediately
deduce the mean-field exponents $\beta$ and $\delta$. The exponent $\beta$ is
determined by $\left.\bar\phi\right|_{h=0}\sim\tau^{\beta}$, which gives $\beta=
1$. The quantity $\delta$ is defined by the relation $\left.\bar\phi \right|_{
\tau=0}\sim h^{1/\delta}$, which implies that $\delta=2$.

To evaluate the exponents $\eta$, $\nu$, and $\gamma$, we need the two-point
inverse correlator given in (\ref{e4}) and (\ref{e5}) for the $\bar\phi=0$ and
the $\bar\phi=2im^2/g$ phases, respectively. The exponent $\eta$ is defined by
$\Delta(q^2)\sim q^{\eta-2}~~(q^2\to\infty)$, which implies that $\eta=0$ in
both phases.

To calculate $\nu$ we need the correlation length. The latter is proportional to
the inverse of the renormalized mass $M$ (the pole of the propagator). In the
mean-field approximation discussed in the present section, $M^2=|m^2|$. For the
phase with nonzero $\bar\phi$ ($\bar\phi=\bar\phi_2$) we find that $\xi^{-2}=
-m^2\sim\bar\phi_2\sim\tau$. This implies that the mean-field exponent $\nu$,
which is defined by $\xi\sim\tau^{-\nu}$, is $\nu=1/2$.

Finally, the exponent $\gamma$ is obtained as follows. The susceptivity $\chi$
is given by $\chi\equiv i\left.\delta\bar\phi/\delta h\right|_{h=0}$. By
differentiating (\ref{e2}) with respect to $h$ we have $m^2\delta\bar\phi/\delta
h+ig\bar\phi\,\delta\bar\phi/\delta h=-i$, which we evaluate at $h=0$ to get
$\left(m^2+ig\bar\phi\right)\chi=1$. We thus obtain
\begin{eqnarray}
\chi&=&(m^2)^{-1}\sim\tau^{-1}\quad{\rm for}~T>T_c~(m^2>0),\nonumber\\
\chi&=&(-m^2)^{-1}\sim\tau^{-1}\quad{\rm for}~T<T_c~(m^2<0).\nonumber
\end{eqnarray}
Therefore, $\gamma$, which is defined by the equation $\chi\sim\tau^{-\gamma}$,
turns out to be $\gamma=1$.

In this and in the previous section we have studied the features of the
continuous transition of the $\cP\cT$-symmetric $ig\phi^3$ theory in the
mean-field approximation by calculating the critical exponents. In the following
section we deepen our understanding of the behavior of the theory in the
critical region. To this end, we compare the conventional and the $\cP
\cT$-symmetric theories by relating their critical behaviors to their stability
properties.

\section{Renormalization and stability}
\label{s4}
In Ref.~\cite{R4} we compared the conventional $\phi^3$ theory with the
corresponding $i\phi^3$ $\cP\cT$-symmetric theory near and at $d=6$ dimensions,
and the instability of the former was contrasted with the stability of the
latter. Moreover, for $d=6-\epsilon$ dimensions, the perturbative
renormalizability of the $\phi^3$ theory (obtained by taking the bare parameters
around the Gaussian fixed point) was compared with the renormalizability of the
$\cP\cT$-symmetric theory. The latter is {\it nonperturbative} because it is
realized around a non-Gaussian fixed point (which collapses onto the Gaussian
fixed point at $d=6$).

As shown in Ref.~\cite{R4}, there is a crucial difference between the RG
properties of the conventional and the $\cP\cT$-symmetric theories. In the
former both $m^2$ and $g$ are relevant directions, but in the latter $m^2$ is
the only relevant direction ($g$ being irrelevant). Accordingly, in
Sec.~\ref{s3} we found that to make the nontrivial solution $\bar\phi_2$ vanish
we only need to tune the temperature $T=2m^2/g$ towards its critical value $T_c=
0$. For generic values of the bare coupling $g$ (which is kept fixed), we must
tune only $m^2$ towards zero. This is similar to what happens in conventional
$\phi^4$ theory, where the only relevant direction is $m^2$ and the quartic
coupling plays no role in determining the phase transition. It appears that the
instability of the $\phi^3$ theory gives rise to an additional condition to
reach the critical region. 

In this section we examine the origin of the profound difference between the two
theories. To this end, we study the mean-field (classical) potential of the
conventional theory
\begin{equation}
V(\phi)=m^2\phi^2/2+g\phi^3/6.
\label{e9}
\end{equation}
As in the previous section, without loss of generality we consider the case
$g>0$.

\begin{figure}[h]
\begin{minipage}{15cm}
\epsfxsize=7cm  
\centerline{\epsffile{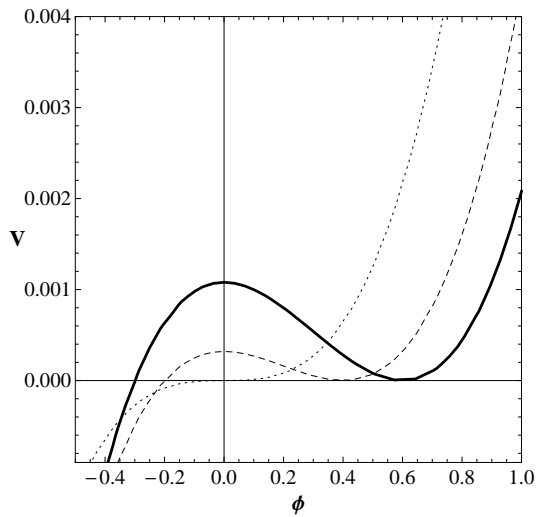}}
\end{minipage}
\caption{The potential $V(\phi)=m^2\phi^2/2+g\phi^3/6-2m^6/(3g^2)$ for $g=6
\times10^{-2}$ and $m^2=-1.8\times10^{-2}$ (solid line), $m^2=-1.2\times10^{-2}$
(dashed line) and $m^2=0$ (dotted line). For the solid and the dashed lines, the
vacuum is at $\bar\phi=-2m^2/g$ while the intersection with the negative $\phi$
axis is at $\phi=m^2/g$. The area $\cA$ of the surface included between the
$\phi$ axis and $V(\phi)$ is proportional to the lifetime of the false vacuum.}
\label{F1}
\end{figure}

This potential is unstable for both the $m^2>0$ and $m^2<0$ cases (see
Fig.~\ref{F1}). However, if the false vacuum is sufficiently long lived (that
is, if the tunneling time is sufficiently long), the theory can be consistently
defined around this vacuum. For $h=0$ the gap equation for the potential in
(\ref{e9}) is $\phi\left(m^2+g\phi/2\right)=0$. For $m^2>0$, the potential has a
minimum at the origin $\bar\phi=0$ and a maximum at $\bar\phi=-2m^2/g$; for
$m^2<0$, there is a minimum at $\bar{\phi}=-2m^2/g$ and the maximum is at $\bar
\phi=0$. Defining the reduced temperature $\tau$ as
\begin{eqnarray}
\tau\equiv m^2/g,
\label{e10}
\end{eqnarray}
we study the continuous phase transition from $\bar\phi=-2m^2/g$ to $\bar\phi=
0$. We stress that in the $\cP\cT$-symmetric $ig\phi^3$ theory (as well as in
the conventional $g\phi^4$ theory) the scaling region $\tau\to0$ is reached by
keeping $g$ finite and taking the limit $m^2\to 0$. This is not true for the
ordinary $\phi^3$ theory.

The potential in (\ref{e9}) for $m^2<0$ is given by the solid line in
Fig.~(\ref{F1}) [apart from a trivial constant shift $V(\phi)\to V(\phi)-2m^6/(3
g^2)$]. In the limit $m^2\to0$ with $g={\rm finite}$ the potential becomes the
dotted line (cubic parabola), which has an inflection point at $\bar \phi=0$. In
this limit the vacuum disappears.

Let us define the area $\cA$,
\begin{eqnarray}
\cA\equiv\int^{-2m^2/g}_{m^2/g} d\phi \,\,V(\phi)=\frac{9m^8}{8g^3},
\label{e11}
\end{eqnarray}
of the surface under the potential function between the points $\bar\phi=-2m^2/
g$ and $\phi=m^2/g$ (see Fig.~\ref{F1}). Note that in the limit $m^2\to0$ with
$g={\rm finite}$ the nontrivial minimum $\bar\phi=-2m^2/g$ moves toward $\bar
\phi=0$, so $\cA$ decreases. Therefore, the lifetime of the false vacuum, which
is proportional to $\cA$, becomes smaller, and the theory is destabilized. As
already mentioned, from the RG viewpoint this happens because in addition to
$m^2$, $g$ is a relevant parameter. To reach the continuous phase transition
($\tau\to0$) while keeping the theory stable ($\cA={\rm fixed}$), we must tune
not only $m^2$ but also $g$. From (\ref{e10}) and (\ref{e11}) we get
\begin{equation}
m^2\sim\cA/\tau^3\qquad{\rm and}\qquad g\sim\cA/\tau^4.
\label{e12}
\end{equation}
This means that in order to reach the continuum limit, $1/m^2$ and $1/g$ must be
tuned separately to zero with $\tau\to0$ according to (\ref{e12}) while keeping
$\cA$ (the vacuum lifetime) fixed. It is clear that the conventional $\phi^3$
theory has an additional relevant direction as compared to its $\cP
\cT$-symmetric counterpart because of the intrinsic instability of the theory.
The $\cP\cT$-symmetric model, being energetically stable, has only one relevant
direction, thus showing a higher degree of predictive power.

\section{Critical exponents near $d=6$}
\label{s5}
In this section we study the critical behavior of the theory for $d<6$. (The
mean-field results provide a good description for $d>6$.) We turn our attention
to the calculation of the critical exponents of the $\cP\cT$-symmetric theory
beyond the mean-field approximation considered in Sec.~\ref{s3}. [According to
the mean-field analysis of Sec.~\ref{s3}, in the presence of the external source
$h$ the two relevant parameters are $\tau=T-T_c=2m^2/g$ (the {\it reduced
temperature}) and $h$ itself (the {\it external field}).] As is known from the
general theory of critical phenomena, below the upper critical dimension ($d=6$
in this case), the fluctuations around the mean-field configuration became
important. RG techniques provide an essential tool for calculating the scaling
behavior of the theory. (Note that although the exponents $\eta$ and $\delta$
appear in the work of Fisher \cite{R8} on the Yang-Lee zero problem for the
first time, the evaluation of the other exponents is an original achievement of
this paper.)

Let us consider our theory in $d=6-\epsilon$ dimensions. The RG equations imply
the existence of two nontrivial fixed points \cite{R4}: 
$$h^{*}=0,\qquad m^{2*}=0,\qquad g^{*}=\pm\sqrt{128\pi^3\epsilon/3}.$$
To each of these points is associated the phase transition that we have just
described in the mean-field approximation. (The above analysis was done for $g^*
>0$ but for $g^*<0$ the results are analogous.)

According to Ref.~\cite{R4} the scalings of $h$, $m^2$, and $g$ with the running
scale $t=\ln(\mu/\mu_0)$ are given by $h(t)=c_1e^{g_1t}$, $m^2(t)=c_2e^{g_2t}$,
and $g(t)=g^*+c_3e^{g_3 t}$, where
\begin{equation}
g_1=-4+4\epsilon/9,\qquad g_2=-2+5\epsilon/9,\qquad g_3=\epsilon.
\label{e13}
\end{equation}
From the above equations we see that the two relevant parameters are $h$ and
$\tau=2m^2/g$.

With the help of the hyperscaling relations
$$\beta=-\frac{d+g_1}{g_2},\qquad\delta=-\frac{g_1}{d +g_1},\qquad
\gamma=\frac{2 g_1+d}{g_2},\qquad\nu=-\frac{1}{g_2},\qquad\eta=2+d+2 g_1,$$
we can calculate from (\ref{e13}) the critical exponents, which turn out to be
\begin{eqnarray}
\beta&=&1,\label{e14}\\
\gamma&=&1+\epsilon/3,\label{e15}\\
\delta&=&2+\epsilon/3,\label{e16}\\
\nu&=&1/2+5\epsilon/36,\label{e17}\\
\eta&=&-\epsilon/9.\label{e18}
\end{eqnarray}
From (\ref{e18}), the scaling dimension of the scalar field is $\left[\phi
\right]=(d-2+\eta)/2=2-5\epsilon/9$. Note that for $\epsilon=0$ these exponents
coincide with the mean-field values calculated in Sec.~III.

\section{Conclusions}
\label{s6}
The $i\phi^3$ model was introduced in \cite{R8} to study the density of the
Lee-Yang zeros of the partition function $Z[H]$ on the imaginary axis of $H$
($H$ being the magnetic field). The ``critical exponents'' in Ref.~\cite{R8} are
not critical exponents in the physical sense, but rather they are parameters
governing the mathematical behavior of the function that gives the asymptotic
density of zeros on the $H$ imaginary axis near the branch point $H=0$. If we
set $m^2=0$ in our theory, we obtain the Lee-Yang model studied in
Ref.~\cite{R8}, where the exponents $\eta$ and $\delta$ already appear; $\beta$,
$\gamma$, and $\nu$ are given for the first time in (\ref{e14}), (\ref{e15}),
and (\ref{e17}).

In contrast to the Lee-Yang model, our $ig\phi^3$ theory has a physical
interpretation as a $\cP\cT$-symmetric Euclidean quantum field theory. Moreover,
to show that our theory undergoes a second-order phase transition at $m^2=0$
(that is, to show the renormalizability of the theory) we have had to
investigate its behavior in the neighborhood of $m^2=0$; it was not sufficient
to study the theory at $m^2=0$. We emphasize that the physical symmetry of the
theory (that is, the $\cP\cT$ symmetry) allows for the presence of the $\phi^2$
operator in addition to the linear and cubic terms.

Using RG techniques, we have studied the theory beyond the mean-field
approximation and have calculated the critical exponents for $d=6-\epsilon$
dimensions up to O$(\epsilon)$. In studying the critical behavior of the $\cP
\cT$-symmetric $ig\phi^3$ quantum field theory, we have shown that the phase
transition is associated with the existence of a nontrivial solution of the gap
equation at a critical value $m^2_c$ of $m^2$. We conclude that one can view the
Lee-Yang model considered in Ref.~\cite{R8} as the critical theory of the $\cP
\cT$-symmetric $ig\phi^3$ model.

Compared with the conventional $\phi^3$ model, the $\cP\cT$-symmetric theory
exhibits new and interesting properties. In particular, it has a higher degree
of predictive power because its critical behavior is governed by one parameter
less than in the $\phi^3$ theory. We have shown that this crucial difference is
related to the different stability properties of the two theories. Thus, from
our work it appears that the renormalization properties of the $\cP
\cT$-symmetric $ig\phi^3$ model, when compared with those of the conventional
$\phi^3$ theory, are quite remarkable and encouraging for further studies and
future applications.

\acknowledgments
CMB thanks the U.S.~Department of Energy and the Leverhulme Foundation and VB
thanks the Istituto Nazionale di Fisica Nucleare (INFN) for financial support.


\begin{thebibliography}{100}
\bibitem{R6} The relevance of these terms is due to the fact that they are the
only parameters governing the behavior of the different physical quantities near
the phase transition. This point is further clarified in Sec.~\ref{s3} with the
computation of the critical exponents.
\bibitem{R7} The gap equation (\ref{e8}) when $h=0$ takes a nicer form when
$T$ is defined as $T=2m^2/g$ rather than through the equivalent definition
$T=m^2$.
\end{thebibliography}
\end{document}